# Collision inside an atom: quantum ion-atom dynamics

Arijit Das[1] *†, Yufeng Wang[1]†, Marcin Welter[2], Karl Horn[1], Dawid Dąbrowski[2], Michał Przybytek[3], Julia Narevicius[1], Michał Tomza[2], Edvardas Narevicius[1]*

[1]Faculty of Physics, Technische Universität Dortmund, Dortmund, Germany.

[2]Faculty of Physics, University of Warsaw, Pasteura 5, 02-093, Warsaw, Poland.

[3]Faculty of Chemistry, University of Warsaw, Pasteura 1, 02-093, Warsaw, Poland

*Corresponding author(s). E-mail(s): arijit2.das@tu-dortmund.de;

edvardas.narevicius@tu-dortmund.de;

Contributing authors: yufeng.wang@tu-dortmund.de;

marcin.welter@fuw.edu.pl; karl.horn@tu-dortmund.de;

dawid.dabrowski@fuw.edu.pl; mitek@chem.uw.edu.pl;

julia.narevicius@tu-dortmund.de; michal.tomza@fuw.edu.pl;

†These authors contributed equally to this work.

## Abstract

Rydberg atoms occupy a central role in quantum science; their loosely bound outer electron gives rise to properties that enable quantum information processing[1] or the formation of exotic molecular bonds[2]. Beyond these roles, they have been proposed as a neutral platform for the observation of quantum effects in collisions and reactions of the ionic core with neutral particles[3,4], a long-standing goal of ion-based chemistry and physics. However, all of the experimental results so far could be explained using a classical description of the internuclear motion[5]. Here we demonstrate cold collisions inside a Rydberg helium atom, achieved by merging helium Rydberg and ground-state helium beams and imaging individual scattering events. We resolve the fully quantum-coherent motion and observe matter-wave diffraction and tunnelling-driven shape resonances down to collision energies of 90 mK. In the case of identical helium nuclei, we demonstrate a new regime of identical-particle Mott interference[6], where quantum statistics begin to govern the total scattering cross section. For distinct isotopes, we show the breakdown of the Born–Oppenheimer approximation and benchmark parameter-free, state-of-the-art theory. We directly measure the helium isotope ionization-energy difference to be $15.0 \pm 0.1$ K. Our experiments establish Rydberg atoms as a platform for the direct observation of quantum dynamics with a general class of ions, atoms and molecules.

## Introduction

A Rydberg atom combines two very different objects, a diffuse excited electron with a tunable orbit size that can reach macroscopic dimensions and an atomic-sized ion core. The orders-of-magnitude difference in the length scales, as shown in Fig. 1a, allows an entire scattering experiment between the ion core and ground-state atom to be placed inside the Rydberg atom, protecting the ion from the detrimental effects of external fields. At low collision energies far from the core, the collision dynamics is set by the interaction of a perturber with the Rydberg electron. This effect was described by Fermi and explains multiple phenomena, from pressure broadening, shifts of Rydberg lines[7], the formation of novel ultra-long-range Rydberg molecules[8] to many-body effects with impurities[9]. Increasing the Rydberg orbit size reduces the Fermi pseudopotential strength and brings the interaction with the ion core to the forefront[3]. This idea was exploited in the study of chemical reactions over a collision energy range spanning several thousand kelvin[10] down to 100 mK[5]; however, the relative motion could always be treated classically. This closely mirrors experiments with bare ions, where strong long-range interactions ensure that many partial waves contribute down to millikelvin energies, making classical trajectory-based descriptions sufficient. Fingerprints of quantum behaviour were detected [11,12,13,14,15,16,17,18] with laser-cooled ions, whilst two experiments demonstrated unambiguous quantum effects in the internuclear motion: Wild *et al.*[19] observed a tunnelling-driven reaction at 15 K and Weckesser *et al.*[20] reported the measurement of Feshbach resonances in collisions between laser cooled Li and $Ba^+$, tuned by an external magnetic field.

Despite the many advantages of ion trapping, such experiments preclude the direct observation of quantum scattering by averaging over many collision events.

Here we image, collision by collision, the cold scattering of $He^+$ off ground-state helium inside a Rydberg helium atom, and find the relative motion to be fully quantum-coherent. Matter-wave diffraction images reveal the shape resonances that govern dynamics at collision energies below 50 K. At the collision energy corresponding to the resonance, particles form a transient molecular bond after tunnelling through the centrifugal barrier as shown in Fig. 1c.

For identical $^4He$ nuclei, the elastic and charge-exchange channels are indistinguishable (Fig. 1b), and their interference introduces a new effect beyond the high-energy Mott oscillations, which only appear as angular fringes in the differential cross section. In the quantum scattering regime where shape resonances dominate, exchange symmetry leads to the cancellation of specific partial waves, leading to changes in the total scattering cross section itself, as seen in Fig. 1d.

For distinct isotopes, the charge-exchange and elastic channels can be distinguished, but the $^3$He–$^4$He ionization-energy difference of 15.0 ± 0.1 K breaks the Born–Oppenheimer (BO) approximation. The endothermic charge exchange between $^3He^+$ and $^4$He is blocked below energies of 15 K, whereas the reverse exothermic process releases this energy into the relative motion, allowing detection of shape resonances at energies as low as 90 mK. We perform the first parameter-free test of beyond-BO theory in a collision-based experiment on a fundamental $He_2^+$ molecular ion, for which the exact beyond-BO theory exists. Our experiments demonstrate a clear path towards observing quantum dynamics in collision and reaction of atoms and molecules that are not amenable to laser cooling and are important players in the chemistry and physics of naturally abundant species.

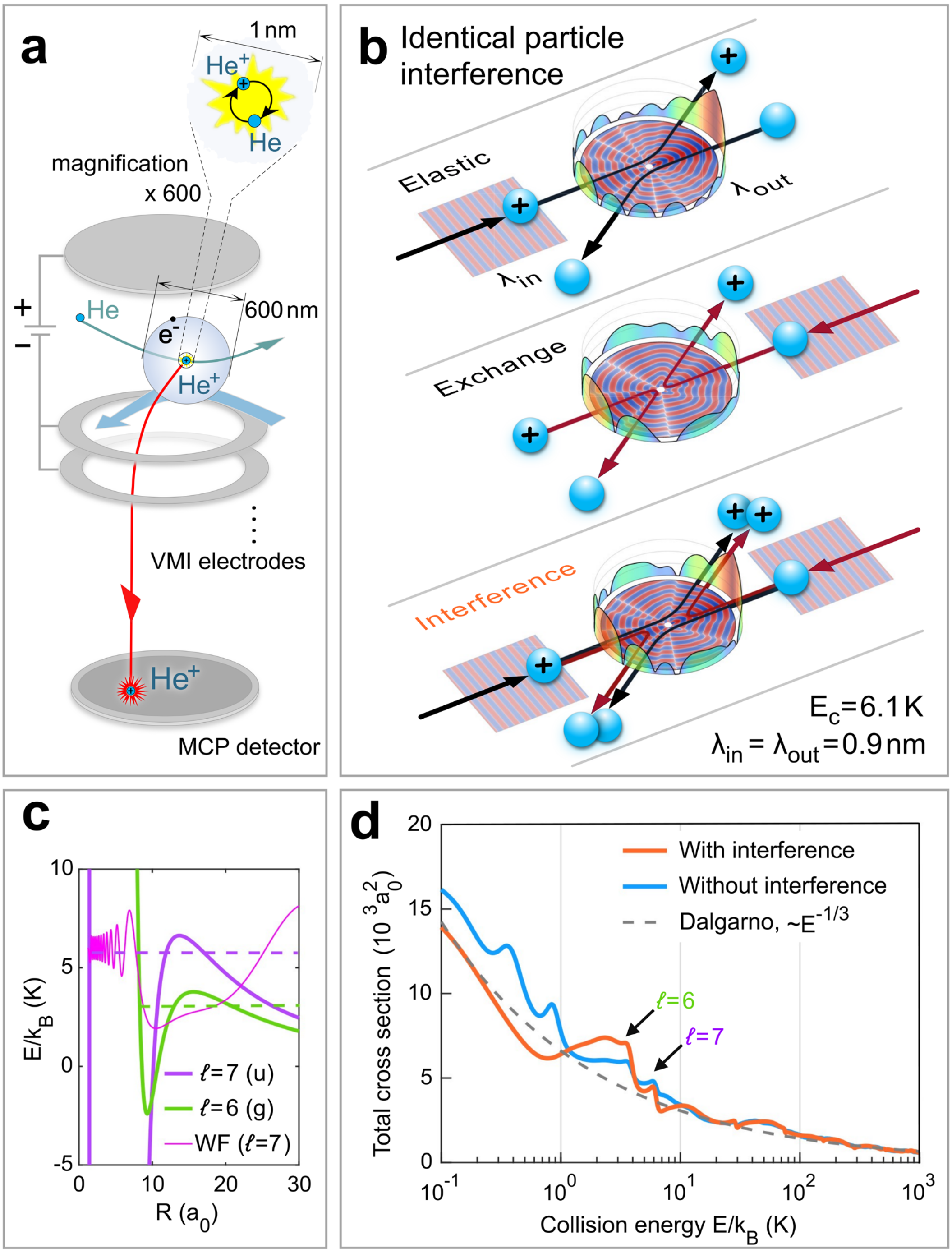

***Fig. 1: Quantum effects in $He^+$–He scattering***. ***a,*** *A schematic experimental setup. A ground-state He atom collides with the ionic core of a Rydberg He atom. The Rydberg electron orbits at a radius far exceeding the ion–atom interaction region; it acts as a spectator, and the collision proceeds as an effectively isolated $He^+$–He collision event. The Rydberg product is subsequently field-ionized by a voltage pulse applied to the VMI electrodes, and the $He^+$ ion is detected on the MCP.* ***b,*** *Theoretical DCSs of the $^4He^+$–$^4He$ at a collision energy of 6.1 K. The striped ribbons represent the incoming (plane-wave) and outgoing de Broglie wavelengths $\lambda_{in/out}$. $^4He^+$ ion is generated through two indistinguishable pathways: elastic scattering (top), in which $^4He^+$ scatters off $^4He$ and remains the same particle, and charge exchange (middle), in which an electron transfers so the incoming neutral leaves as $^4He^+$. Because the two $^4He$ nuclei are identical, the amplitudes add coherently (bottom) which give rise to interference.* ***c,*** *$^4He_2^+$ potential-energy curves (PECs) with centrifugally modified $\ell = 6$ and $\ell = 7$ wells that each support a quasi-bound resonance state (energies marked by dashed lines); the radial wavefunction (WF) of the resonance in the $\ell = 7$ well is overlaid. Labels g and u denote the $A^2\Sigma g^+$ and $X^2\Sigma u^+$ electronic states of $^4He_2^+$, respectively.* ***d,*** *The integral cross section as a function of collision energy, evaluated with and without the interference term and compared with the classical Dalgarno result ($\propto E$^(−1/3))[21]. The two labelled features arise from $\ell = 6$ and $\ell = 7$ partial-wave shape resonances.*

**Homonuclear $^4He^+/^4He$ case: Identical-particle interference, shape resonances**

To study helium atom and ion collisions, we merge two beams, a ground-state He ($1^1S_0$) beam on a straight path, and a metastable He ($2^3S_1$) beam, produced by dielectric barrier discharge and steered by a curved magnetic guide to reach low collision energies in the moving frame of reference. After merging, the metastable atoms are excited to a Rydberg state (n = 70) by 260 nm laser radiation and continue into the velocity map imaging (VMI) detector where they collide with ground-state helium (see Methods section for details). Both the Rydberg collision products and the uncollided Rydberg atoms are then field-ionized. The ions are imaged in a VMI setup with a microchannel plate (MCP) detector and phosphor screen (Fig. 1a).

At high principal quantum numbers, such as n = 70, the Rydberg electron behaves essentially as a spectator. We estimate that only below n = 27 does its influence become significant (see Methods). There, following the Fermi pseudopotential model, the electron-induced energy shift becomes on the order of the *p*-wave barrier height and the collision can no longer be treated as a pure ion–atom process. Our experiment operates far above this cut-off with the measured dynamics reflecting the bare $He^+$ + He potential.

In the homonuclear $^4He^+$ + $^4He$ system, exchange symmetry between the identical nuclei requires the electronic wavefunction to be either symmetric (gerade, $A^2\Sigma_g^+$) or antisymmetric (ungerade, $X^2\Sigma_u^+$) under inversion, giving rise to both a shallow (~25 K) and a deep potential curve (~28710 K), respectively (Extended Data Fig. 1). We obtained the potential energy curves (PECs) using full configuration-interaction calculations, matched with the long-range part having the leading $1/R^4$ and several higher-order terms[22]. A charge initially localized on one nucleus is a superposition of the two electronic states mentioned above. During the collision a phase difference accumulates between the gerade and ungerade states. Asymptotically this phase fixes the amplitude for finding the charge on either nucleus, and thus the ratio between the elastic and charge-exchange channels. The elastic and charge-exchange amplitudes are, respectively, the sum and difference of the gerade and ungerade scattering amplitudes,

$$f_{el} = \tfrac{1}{2} ( f_g + f_u ), \qquad f_{ce} = \tfrac{1}{2} ( f_g - f_u ). \qquad (1)$$

Note that, for the resonant charge exchange there is no coupling between the two curves and the collision outcome can be described within the BO framework.

The scattering dynamics are therefore governed by two indistinguishable processes, elastic scattering and charge exchange. As shown in Fig. 1b, a $^4He^+$ ion elastically scattered through an angle θ cannot be distinguished from a $^4He^+$ ion scattered into the $\pi - \theta$ direction after its neutral partner has undergone charge exchange. The two amplitudes must therefore be added coherently, giving rise to an interference term in the DCS,

$$I_{total}(\theta) = | f_{el}(\theta) + f_{ce}(\pi - \theta) |^2 \qquad (2)$$

Expanding this expression separates the individual elastic and charge-exchange contributions from the interference term,

$$I_{total}(\theta) = | f_{el}(\theta) |^2 + | f_{ce}(\pi - \theta) |^2 + 2\,\mathrm{Re}\{ f_{el}(\theta)\, f_{ce}^*(\pi - \theta) \} \qquad (3)$$

Experimentally, we obtain velocity map images at several different collision energies, representative examples of which are shown in the insets of Fig. 2. Atoms that did not collide cannot be separated from the scattered ones and also appear on the VMI detector, masking the low-scattering-angle part of the collision sphere in velocity space. We observe pronounced diffraction oscillations in the images, analogous to matter-wave diffraction, which arise from interference between different partial waves, an effect previously observed only in neutral–neutral collisions[23,24]. We obtained the partial rate coefficients (Fig. 2) by counting the number of $He^+$ ions in the backward-scattering hemisphere (144–180°) of images down to 0.8 K. We observe several peaks in the experimental partial rate coefficients, which are direct evidence of shape resonances governing the collisional dynamics.

We compute the theoretical rate coefficient using Eq. (3) (orange curve in Fig. 2). Reproducing both the resonance positions and the correct amplitude of the rate coefficient requires the full symmetrized treatment. Conversely, a calculation that omits the interference term (blue curve in Fig. 2) deviates markedly from the experimentally obtained reaction rate. In many cases, the interference term has been neglected, which is a valid approximation for high-energy collisions[25].

This interference effect is fundamentally distinct from the Mott oscillations reported previously for $He^+$ + He scattering at high collision energies[26]. Mott scattering, the interference between the two indistinguishable outgoing channels related by exchange of the ion and the atom, manifests solely in the differential (angular) cross section. Arising from the interference of classical scattering paths, it has to date been observed only at collision energies above 100 eV[26], where the integrated cross sections computed with and without the interference term are nearly identical, since the rapid angular oscillations average out upon integration over the scattering angle. The interference term is therefore not needed to reproduce the measured reaction rate in that regime.

By contrast, the effect reported here persists in the integrated cross section up to collision energies on the order of 10 K. At low collision energy, a single partial wave contribution becomes dominant, leading to a pronounced interference effect in the total cross section as well as the appearance of shape resonances.

In the $^{4}He^{+} + {}^{4}He$ system we resolve five resonance peaks, at (1.53 ± 0.06) K, (3.2 ± 0.1) K, (6.1 ± 0.1) K, (16.6 ± 0.2) K, and (31.5 ± 0.2) K (uncertainties represent one standard deviation). By performing partial wave analysis, we assign these peaks to the partial waves $\ell = 4$ (gerade), $\ell = 6$ (gerade), $\ell = 7$ (ungerade), $\ell = 8$ (gerade), and $\ell = 11$ (ungerade), respectively. The systematic alternation, gerade electronic symmetry for the even-$\ell$ resonances and ungerade for the odd-$\ell$ ones, is enforced by the exchange symmetry of the two identical spin-zero $^{4}He$ nuclei. Bose statistics require the total wavefunction to be symmetric under nuclear exchange, tying the g/u parity of the electronic state to the rotational parity (−1)^ℓ and permitting only even-$\ell$ gerade and odd-$\ell$ ungerade combinations. The observed series is thus a direct fingerprint of this indistinguishability. On account of their sharpness on a few-kelvin scale, these shape (orbiting) resonances, which represent quasi-bound states trapped behind the centrifugal barrier of the long-range polarization potential, reflect the small number of contributing partial waves, the same condition that allows the interference to survive integration.

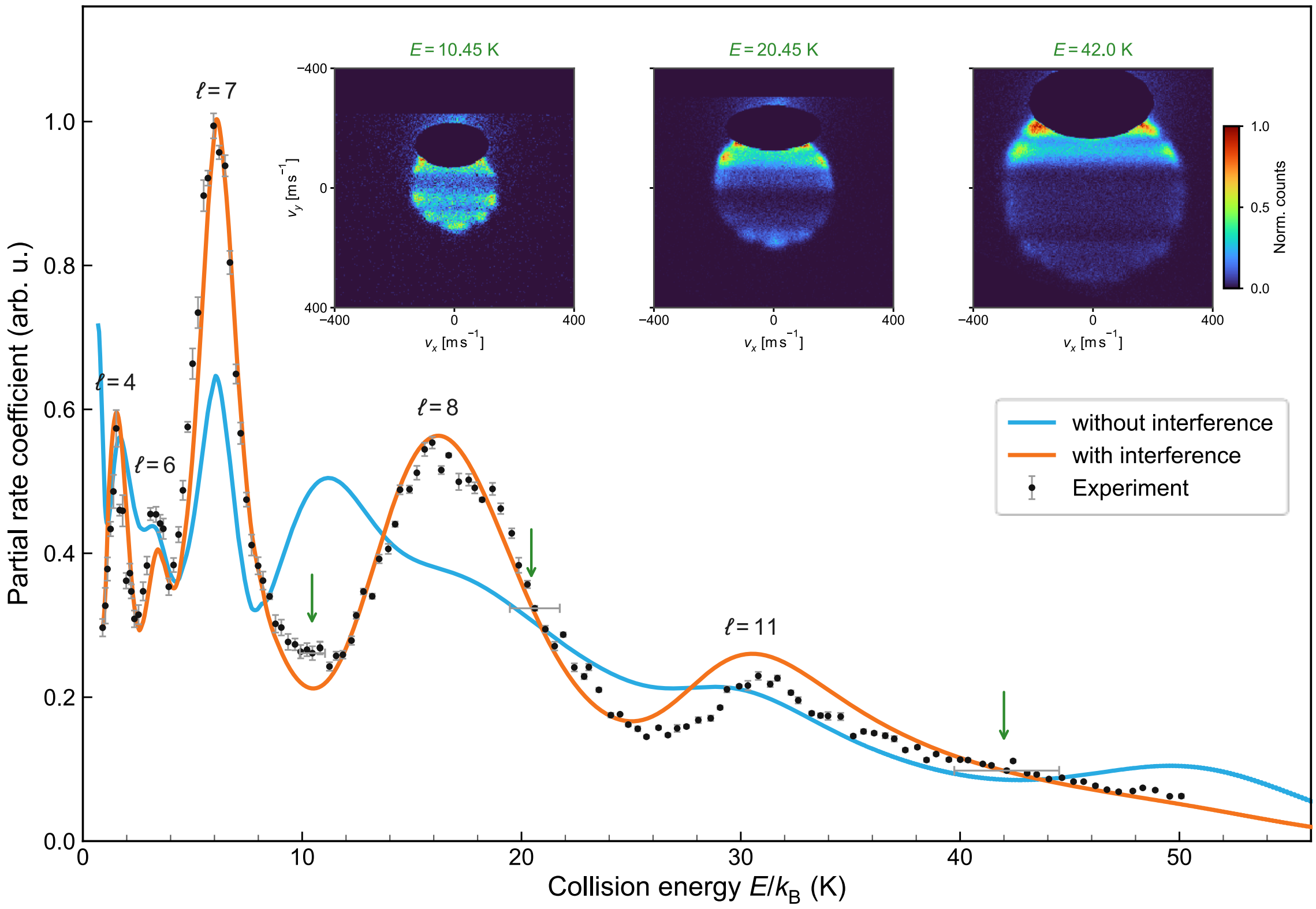

***Fig. 2: Partial reaction rate coefficient and velocity mapped images of $^4He^+$–$^4He$ collision.*** *Normalized partial rate coefficient as a function of collision energy. Black circles show the experimental data, horizontal error bars on green arrows denote the spread in collision energy, and vertical error bars the statistical uncertainty of the measured rate. The orange and blue curves are the theoretical rate coefficients calculated with and without inclusion of quantum interference contributions, respectively. Pronounced maxima are assigned to shape resonances of orbital angular momenta* $\ell$ = *4, 6, 7, 8 and 11, as labelled. Green arrows mark the three collision energies* ($E/k_B$ = *10.45, 20.45 and 42.0 K) at which the velocity map images shown in the insets were recorded. The insets display the corresponding product-ion velocity distributions in the* $v_x$ *and* $v_y$ *plane; the colour scale gives the normalized counts. The y-axis is the direction of the relative velocity vector, with the forward direction* ($\theta = 0°$) *pointing up in all of the images.*

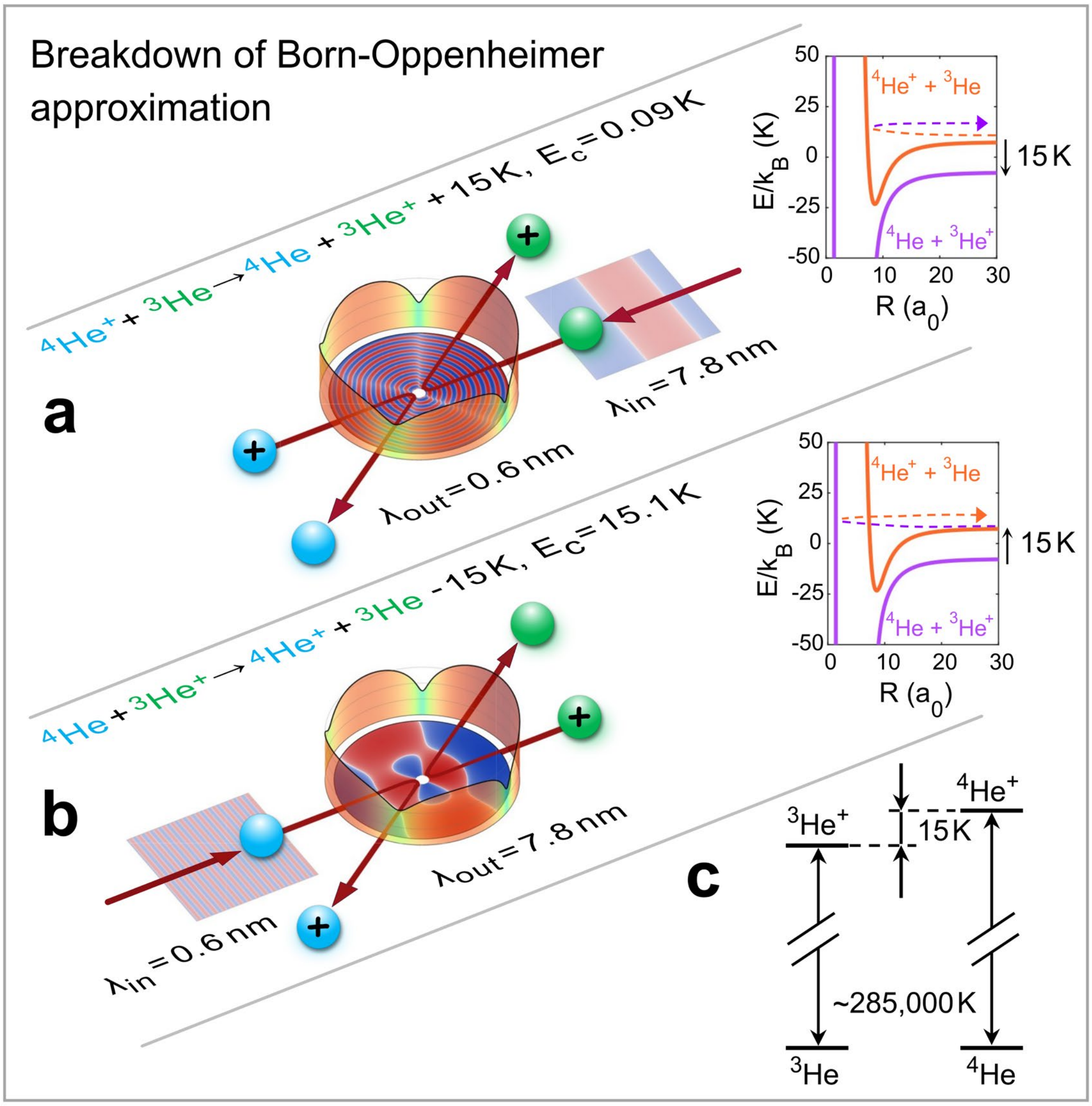

*__Fig. 3: Beyond–Born–Oppenheimer (BO) effects in charge exchange of a heteronuclear ion–atom system. a,__ Exothermic direction, $^4He^+ + {}^3He \rightarrow {}^4He + {}^3He^+$ (energy released, +15 K; collision energy $E_c = 0.09$ K). The theoretical DCS is rendered as a circular disk with interference fringes. The incoming and outgoing de Broglie wavelengths are markedly unequal ($\lambda_{in}$ = 7.8 nm, $\lambda_{out}$ = 0.6 nm). The associated PECs (right) show the two diabatic states — $^4He^+ + {}^3He$ (orange) and $^4He + {}^3He^+$ (purple) — separated by the ≈15 K isotope defect; the downward arrow marks the energy released as the system moves to the lower product curve. __b,__ Endothermic direction, $^4He + {}^3He^+ \rightarrow {}^4He^+ + {}^3He$ (energy absorbed, −15 K; $E_c$ = 15.1 K). Its DCS is shown as in __a__, with the wavelengths reversed ($\lambda_{in}$ = 0.6 nm, $\lambda_{out}$ = 7.8 nm). In the corresponding PECs (right) the upward arrow marks the ≈15 K that must be supplied to reach the higher product asymptote. __c__, Energy-level scheme. The ionization energies of neutral $^3He$ and $^4He$ each lie ≈285,000 K (≈24.6 eV) above their ground-states, yet the $^3He^+$ and $^4He^+$ product asymptotes are offset by ≈15 K. This tiny isotope shift arises from the finite nuclear mass, a beyond–BO mass-dependent correction to the electronic energy, and is what breaks the symmetry between the two charge-exchange directions.*

**Heteronuclear $^4He^+/^3He$ and $^3He^+/^4He$ case: Beyond BO, shape resonances and reaction threshold**

In contrast, for the heteronuclear systems $^4He^+ + {}^3He$ and $^3He^+ + {}^4He$ (Fig. 3), the isotopic mass difference breaks the exchange symmetry, so that the two channels, elastic scattering and charge-exchange, become experimentally separable. We focus on the charge-exchange channel, where time-of-flight mass selection isolates the charge-exchange products and recovers the full angular distribution of the scattering image. Because charge exchange is no longer resonant in these systems, the dynamics cannot be captured within the BO framework and require the inclusion of non-adiabatic couplings; this is the general case of the charge-exchange process. The two directions of the reaction differ in energy by 15 K, set by the difference in ionization potential between $^4He$ and $^3He$, which sits atop the ~285,000 K ionization energy of the neutral atoms (Fig. 3c).

For $^4He^+ + {}^3He$, the charge-exchange process is exothermic, releasing 15 K of electronic energy into the relative nuclear motion. As a result, even at the lowest collision energies, the size of the scattering sphere is set by the reaction exothermicity rather than by the collision energy, giving ready access to millikelvin collision energies with fully resolved angular structure. This is illustrated in the velocity map image measured at 0.09 K, where the $\ell = 2$ partial wave dominates (Fig. 4a, inset). The slow incoming collision has a de Broglie wavelength of 7.8 nm, while the products, accelerated to ≈15.1 K by the released energy, leave with a much shorter wavelength of 0.6 nm (Fig. 3a). The rate coefficient in Fig. 4a reveals three resonances, at (90 ± 11) mK, (2.5 ± 0.1) K, and (6.5 ± 0.4) K, which we attribute to the $\ell = 2$, $\ell = 5$, and $\ell = 6$ partial waves, respectively. Reproducing these measurements theoretically, in particular the resonance positions, requires a treatment beyond BO (Fig. 4a, orange curve). Neglecting the non-adiabatic terms fails to describe the data at low collision energy, and the classical Langevin rate underestimates the measured rate by a factor of two in the same range.

The reverse channel, $^3He^+ + {}^4He \rightarrow {}^4He^+ + {}^3He$, is the mirror image of this process and is endothermic: a minimum collision energy of about 15 K is required to overcome the ionization-potential difference and open the channel. The roles of the incoming and outgoing waves are therefore exchanged, a fast collision at 15.1 K (de Broglie wavelength 0.6 nm) gives up 15 K to leave slow products (7.8 nm) (Fig. 3b), and the size of the scattering sphere is now governed by the endothermicity of the reaction, as shown by the velocity map image measured at 26 K in the inset of Fig. 4b. The total rate coefficient (Fig. 4b) exhibits a clear, well-defined threshold.

Fitting the measurements to a convoluted theoretical cross section with the threshold as the only free parameter yields 15.0 ± 0.1 K, in excellent agreement with the value of 15.11 ± 0.07 K from the early spectroscopic work of Herzberg[27] and with the most accurate *ab initio* result, 15.0466 ± 0.0001 K[28]. We note that our threshold measurement is statistically dominated by $^3He^+$ ions in the F = 1 hyperfine level. Because $^3He$ has a negative nuclear magnetic moment, the ground-state hyperfine structure of $^3He^+$ is inverted, placing F = 1 below the hyperfine centre by one quarter of the total splitting of 0.416 K. The expected threshold is therefore shifted to 15.0466 + 0.416/4 K = 15.15 K.

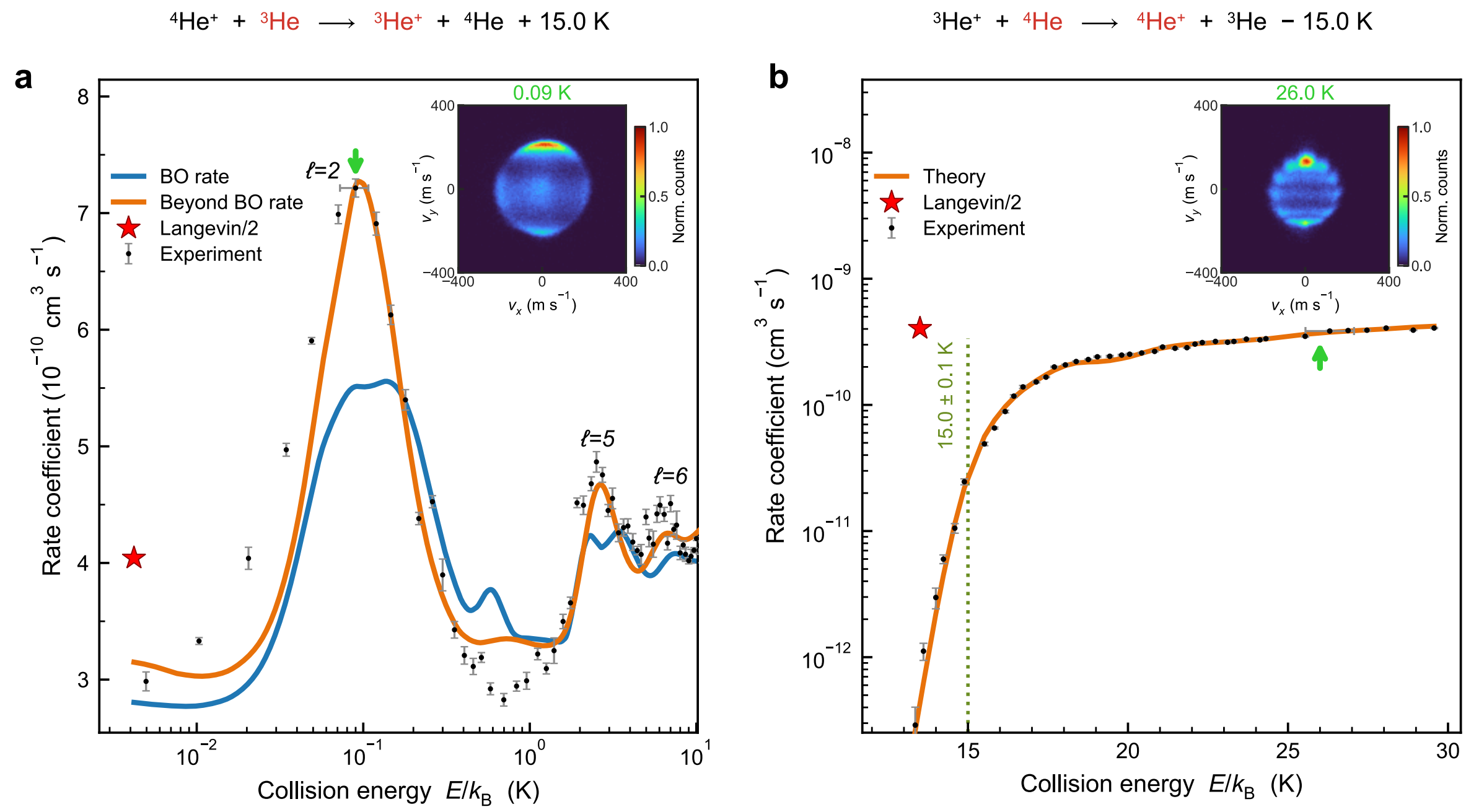

*Fig. 4: Charge-exchange rate coefficients for the isotope-exchange system and their velocity map images.* ***a****, Rate coefficient for the exothermic forward reaction $^4He^+ + {}^3He \rightarrow {}^3He^+ + {}^4He$ (+15.0 K) as a function of the collision energy. Black circles show the experimental data; the orange and blue curves are the theoretical rate coefficients calculated with ("Beyond BO rate") and without ("BO rate") non-adiabatic couplings, respectively. Pronounced maxima arising from shape resonances in the incident partial waves ℓ = 2, 5 and 6 are indicated.* ***b****, Rate coefficient for the endothermic reverse reaction $^3He^+ + {}^4He \rightarrow {}^4He^+ + {}^3He$ (−15.0 K), shown on a logarithmic scale; black circles are the experimental data and the orange curve is theory. The olive dotted line marks the reaction threshold at 15.0 ± 0.1 K, in agreement with the known reaction endothermicity. In both panels the red star indicates half the energy-independent Langevin capture rate. Vertical error bars denote statistical uncertainties; horizontal error bars represent the collision-energy spread and are shown only at the point indicated by the green arrow in each panel. For both panels, the experimental data reporting relative rates is multiplied by a factor to match the theoretically obtained rate coefficients. Insets, normalized velocity map images of the product ions ($^3He^+$ in* ***a*** *and $^4He^+$ in* ***b****) recorded at the collision energies indicated by the green arrows ($E/k_B$ = 0.09 K in* ***a*** *and 26.0 K in* ***b****); axes are the in-plane velocity components $v_x$ and $v_y$, and the colour scale gives the normalized ion count. The y-axis is the direction of the relative velocity vector, with the forward direction (θ = 0°) pointing up in all of the images.*

## Conclusions and future perspectives

In this work, we demonstrate quantum effects in ion-atom collisions by observing matter-wave diffraction, a new interference effect in identical-particle scattering that is observable in the integrated cross section, and shape resonances in ion-atom collisions at collision energies as low as 90 mK in the heteronuclear case, where the BO-approximation is no longer valid. Our present study focuses on a fundamental atomic system; the same approach can be easily extended to molecular systems. Both reactions with neutral molecules as well as molecular ion reactions with neutral species can be studied with high resolution at low collision energies. Molecules can be excited to a specific rovibrational Rydberg state, enabling quantum state-to-state studies of ion-neutral reactions with velocity map imaging providing the tomographic information on the final quantum state distribution. We expect that all the aspects of quantum scattering will be important in these interactions, with shape resonances, Feshbach resonances and identical particle scattering effects shaping the dynamics in systems that are not amenable to laser cooling.

## Acknowledgements

E.N. acknowledges support from the Cluster of Excellence RESOLV and Alexander von Humboldt Professorship endowed by the Federal Ministry of Research, Technology and Space. We thank Christiane Koch for her useful comments on the manuscript.

**Funding:** This work was supported by ERC Advanced Grant No.885180-MOLBEC.

We gratefully acknowledge the National Science Centre Poland (grant no. 2020/38/E/ST2/00564) for financial support and the Poland’s high-performance computing infrastructure PLGrid (HPC Center: ACK Cyfronet AGH) for providing computer facilities and support (computational grant no. PLG/2025/018722).

**Author contributions:** AD, YW, JN, and EN conceived and conducted the experiments. MW, KH, DD, MP, and MT performed theoretical calculations. All authors contributed to writing and revising the manuscript.

**Competing interests:** The authors declare no competing interests.

## Methods

## Experimental Method

**Supersonic beam generation and characterization:** We investigated three isotopic combinations using a merged molecular beam setup: $^4He^+ + ^4He$, $^4He^+ + ^3He$, and $^3He^+ + ^4He$. The technical details of the apparatus can be found elsewhere[29].

The experiment utilizes two Even-Lavie valves[30]. For the $^4He^+ + ^4He$ experiments, a beam of helium atoms was produced with 99.9999% pure helium at a backing pressure of 20 bar. The helium atoms are subsequently excited to the metastable state (He*) using a dielectric-barrier discharge[31]. With the valve temperature maintained at 75 K, the metastable beam was produced at a mean velocity of 927 ± 16 m/s. The other Even-Lavie valve prepares a helium beam in the ground-state traveling directly into the collision chamber. Two skimmers, each with a 4 mm diameter orifice, are positioned 10 cm downstream of the valve nozzle. The He* beam is then deflected by a curved magnetic dipole. Only the low-field-seeking Zeeman sublevel is guided towards the collision volume during the transit. At the exit of the guide, the deflected He* beam is merged with a straight ground-state helium beam. Two collimating apertures—a 3 mm diameter circular orifice and a 3 × 2 mm elliptical orifice—are placed 440 mm from the valve in order to collimate the beams. The distance between the apertures is 17.5 mm. The collimated metastable helium beam is then excited to the Rydberg state just before entering the collision volume. Excitation to the Rydberg level (n=70) was achieved using a direct transition at 260 nm light produced by a tripled pulsed dye laser (LDS 798 dye) pumped by a pulsed (10 Hz) Nd:YAG laser operating at 532 nm. The typical pulse energy at 260 nm was 0.8–1.2 mJ. The velocity of the Rydberg beam was characterized by monitoring the ion signal as a function of the laser excitation delay. Our merged-beam setup enables scanning of the collision energy in the low-energy regime with high resolution by exploiting the phase-space correlation that develops during beam propagation from the source to the detection region[32]. The collision energy was determined by the mean relative velocity between the beams. The ground-state helium beam valve temperature has been varied between 94 K and 226 K, yielding a collision energy range of 0.8 K to 50.0 K. The ground-state helium beam was characterized by a time-of-flight mass spectrometer (TOF-MS), which contains an electron impact ionization element. The TOF–MS was operated in a multi-pulse mode with 7.3 μs intervals to obtain the longitudinal profile and the velocity of the ground-state helium beam.

For the $^4He^+ + ^3He$ system, a $^4He$ Rydberg beam was generated using the same conditions as described above. We utilized 99.9% pure $^3He$ at a backing pressure of 15 bar to generate the $^3He$ ground-state beam. By tuning the $^3He$ valve temperature from 63 K to 107 K, we accessed the near s-wave scattering regime, covering collision energies from 5 mK to 10 K.

For the $^3He^+ + ^4He$ system, $^3He$ Rydberg atoms were generated from a metastable beam prepared from a discharge source at a backing pressure of 15 bar and subsequently excited by a laser to the n=76 Rydberg state. The $^3He$ valve temperature was fixed at 55.5 K, corresponding to a mean velocity of 927 ± 26 m/s, while the $^4He$ beam temperature was varied from 150 K to 202 K. This resulted in collision energies ranging from 12.7 K to 30 K.

**Velocity Map Imaging (VMI) Detection:** The Rydberg helium atoms enter the VMI detection region under field-free conditions. The apparatus employs a double-VMI configuration[33] capable of detecting both ions (top) and electrons (bottom); however, in the present experiment, only the ion-side MCP and camera were used. The VMI stack consists of eight electrodes on the bottom and eleven on the top side. In this work, twelve plates were driven by pulsed voltages with a rising edge of 30 nsec, while the remaining electrodes were maintained at constant voltages. These constant-voltage electrodes are located 90 mm from the VMI center, and SIMION simulations verified that the electric fields they generate within the collision volume are negligible.

The pulsed voltages were switched on such that they coincided with the arrival of the Rydberg helium atoms at the center of the VMI. By adjusting the relative delay between the two beams, spatial and temporal overlap of the beams at the center of the VMI was ensured. Upon pulsing, both scattered and unscattered Rydberg atoms were field-ionized, and the resulting helium ions were accelerated toward the MCP detector. The MCP detector used in the detection region has an outer diameter (OD) of 50 mm with an active area of 46 mm. The detector consists of two MCPs in a chevron configuration followed by a P43 phosphor screen. The fluorescence generated from the electrons (formed from $He^+$ hitting the MCP) as they impinge on the phosphor screen is then imaged on the camera. A real-time centroiding algorithm determines the x–y coordinate of each event on the MCP, correlating an electrical signal with the position and number of ions.

Standard operating voltages for the first two VMI stack electrodes were set to 558 V and 506 V, respectively, and decreased to 0 V by the last one. The distance between the first and the second electrode is 20 mm, resulting in an electric field of approximately 26 V/cm for field ionization. In order to increase the resolution, the flight distance from the VMI center to the MCP is 1.75 m. The TOF duration was 12.27 µs for $^4He^+$ ions and 10.13 µs for $^3He^+$ ions. A fine-grid nickel mesh (SPI Supplies, 500 lines per inch, hole size 39 µm, 60% open area) is placed 7 mm in front of the MCP detector and kept grounded. The front MCP was biased at −1245 V to ensure that the helium ions possessed sufficient kinetic energy for optimal detection efficiency. The presence of the mesh allows operation at lower VMI voltages and a longer TOF, enlarges the scattering Newton spheres while enabling faster pulsing. A 200 ns gating pulse was applied to the front MCP to selectively detect the $^4He^+$ and $^3He^+$ ions. For the homonuclear case $^4He^+ + {}^4He$, a large background arises from the unscattered Rydberg $^4He$ beam. To protect the detector from the intense primary ion beam and to prevent interference from secondary-electron repulsion in the MCP detector, a 1.4 mm-wide stainless-steel pin was positioned approximately 25 mm in front of the MCP detector to block the primary beam. This pin was mounted on an XY-axis manipulator for precise positioning. At low collision energies, where the Newton sphere is small, the number of Rydberg atoms was intentionally reduced to ensure accurate imaging.

The VMI calibration was performed by varying the laser delay to ionize different velocity groups of the metastable helium beam inside the VMI chamber. By changing the delay, different portions of the longitudinal velocity distribution were selectively ionized, and the resulting shift in the beam position in the velocity-map images was measured. For this purpose, the laser delay was varied over a 70 µs time window, corresponding to metastable helium velocities ranging from 858 to 974 m/s. The velocity-per-camera pixel was determined to be 1.89 m/s for the $^4He^+$ ions. The experimental VMI images shown in Fig. 2 and Fig. 4 were obtained by binning the x–y coordinates of each MCP event onto a 2D array with a mesh spacing equal to 2 camera pixels.

**Reaction Rate Measurements:** To determine the reaction rates shown in Fig. 2 and Fig. 4, the scattering signals for the $He^+ + He$ system were measured as a function of collision energy. Collision energies were scanned by keeping the Rydberg beam velocity constant, while adjusting the valve temperature to vary the ground-state helium beam velocity, which was monitored via TOF-MS. We accumulated the scattering signal over 12,000 shots to achieve a reasonable signal-to-noise ratio. For the $^4He^+ + {}^4He$ case, we counted the ion numbers of the backward-scattering sphere from an angle of 144° to 180°. For the $^4He^+ + {}^3He$ and $^3He^+ + {}^4He$ cases, we counted the ion numbers of the whole scattering sphere. The reaction rate was obtained by measuring the scattering signal and normalizing it, taking both the ground-state helium beam, characterized via TOF-MS, and the Rydberg helium beam, monitored by counting ions in the imaging detector, into account. The error bar in the measurements primarily comes from the fluctuation of laser power with time. The error was significantly reduced by monitoring the laser intensity with a photodiode (DET10A2, Thorlabs Inc.), finally correlating the amplitude of the signal from the photodiode to the number of ions measured by the imaging detector.

**Electron spectator role:** A high-n helium Rydberg electron (the spectator) couples to a nearby polarizable perturber through an s-wave contact (Fermi) term set by the electron–perturber scattering length $a_s$, while the $He^+$ core attracts that same perturber via the long-range polarization potential $-C_4/r^4$. The perturber's relative motion sits at the top of its lowest centrifugal (p-wave, ℓ = 1) barrier, whose height fixes the energy E. Then $n_{max}$ is the largest principal quantum number for which the Fermi shift can still match that barrier energy, ΔE = E.

$$n_{max} = 2^{5/6}\left(\frac{a_s}{2\sqrt{2}\pi}\right)^{\frac{1}{3}} C_4{}^{1/12}\mu^{5/12}$$

For this p-wave barrier (ℓ = 1), $n_{max} = 27$ for the $^4He^+/^4He$ collision, above which the interaction can be treated as a pure ion–atom collision.

## Theoretical Methods

Derivation of the coupled-channel Hamiltonian

**General heteronuclear framework:** Consider a diatomic molecule consisting of two nuclei with masses $M_A$ and $M_B$ located at space-fixed positions $\mathbf{R}_A$ and $\mathbf{R}_B$, respectively, and $N$ electrons at space-fixed positions $\mathbf{r}_i$. The total non-relativistic Hamiltonian for this system has the form

$$\hat{H} = -\frac{1}{2M_A}\nabla^2_{\mathbf{R}_A} - \frac{1}{2M_B}\nabla^2_{\mathbf{R}_B} - \frac{1}{2}\sum_{i=1}^{N}\nabla^2_{\mathbf{r}_i} + \hat{V}, \qquad \text{(S1)}$$

where the potential energy operator $\hat{V}$ depends only on the interparticle distances. Switching to the body-fixed frame related to the center of nuclear mass (CNM) and neglecting the CNM motion, the Hamiltonian from Eq. (S1) becomes

$$\hat{H} = -\frac{1}{2\mu}\nabla^2_{\mathbf{R}} - \frac{1}{2M}\nabla^2_{\mathrm{e}} + \hat{H}_{\mathrm{el}}, \qquad \text{(S2)}$$

where $M = M_A + M_B$, $\mu = M_A M_B / M$ is the reduced mass of the nuclei, $\mathbf{R} = \mathbf{R}_A - \mathbf{R}_B$ is the vector connecting the two nuclei, $\nabla_{\mathrm{e}}$ is defined as

$$\nabla_{\mathrm{e}} = \sum_{i=1}^{N}\nabla_{\mathbf{r}_i}, \qquad \text{(S3)}$$

and $\hat{H}_{\mathrm{el}}$ is the standard clamped-nuclei electronic Hamiltonian,

$$\hat{H}_{\mathrm{el}} = -\frac{1}{2}\sum_{i=1}^{N}\nabla^2_{\mathbf{r}_i} + \hat{V}. \qquad \text{(S4)}$$

The solutions to the clamped-nuclei, or Born–Oppenheimer (BO), problem, $\psi_\alpha(\mathbf{r}_i;R)$, depend on $R$ (the length of the vector $\mathbf{R}$) only parametrically and are orthonormal, i.e.,

$$\langle\alpha|\beta\rangle = \int \psi_\alpha(\mathbf{r}_i;R)^*\,\psi_\beta(\mathbf{r}_i;R)\,\mathrm{d}\tau = \delta_{\alpha\beta}, \qquad \text{(S5)}$$

where $\mathrm{d}\tau$ represents integration over all electronic coordinates. By assuming that the BO functions are real, and by differentiating the orthogonality condition with respect to the coordinates of the internuclear vector, we arrive at two useful relations:

$$\langle\alpha|\nabla_{\mathbf{R}}\beta\rangle = \begin{cases} 0, & \alpha = \beta, \\ -\langle\beta|\nabla_{\mathbf{R}}\alpha\rangle, & \alpha \neq \beta, \end{cases} \qquad \text{(S6)}$$

and

$$\langle\alpha|\nabla^2_{\mathbf{R}}\beta\rangle = \begin{cases} -\langle\nabla_{\mathbf{R}}\alpha|\nabla_{\mathbf{R}}\beta\rangle, & \alpha = \beta, \\ -\langle\nabla_{\mathbf{R}}\alpha|\nabla_{\mathbf{R}}\beta\rangle + \nabla_{\mathbf{R}}\langle\alpha|\nabla_{\mathbf{R}}\beta\rangle, & \alpha \neq \beta. \end{cases} \qquad \text{(S7)}$$

In the electronic two-state model, a solution to the Schrödinger equation with the Hamiltonian from Eq. (S2) is sought in the form

$$\Psi(\mathbf{r}_i,\mathbf{R}) = f_1(\mathbf{R})\,\psi_1(\mathbf{r}_i;R) + f_2(\mathbf{R})\,\psi_2(\mathbf{r}_i;R), \qquad \text{(S8)}$$

where $f_\alpha(\mathbf{R})$, $\alpha = 1,2$, are the nuclear functions to be determined. After inserting Eq. (S8) into the Schrödinger equation, multiplying from the left by $\psi_\alpha(\mathbf{r}_i;R)^*$, and integrating over the electronic coordinates, we arrive at the nuclear problem governed by a $2\times 2$ Hamiltonian matrix operator,

$$\mathbf{H}(R) = \begin{pmatrix} -\frac{1}{2\mu}\nabla^2_{\mathbf{R}} + D_{11} + V_1 & -\frac{1}{2}(\mathbf{F}_{12}\cdot\nabla_{\mathbf{R}} + \nabla_{\mathbf{R}}\cdot\mathbf{F}_{12}) + D_{12} \\ \frac{1}{2}(\mathbf{F}_{12}\cdot\nabla_{\mathbf{R}} + \nabla_{\mathbf{R}}\cdot\mathbf{F}_{12}) + D_{12} & -\frac{1}{2\mu}\nabla^2_{\mathbf{R}} + D_{22} + V_2 \end{pmatrix}, \qquad \text{(S9)}$$

where $D_{\alpha\beta}$ are the symmetric matrix elements,

$$D_{\alpha\beta} = \frac{1}{2\mu}\langle\nabla_{\mathbf{R}}\alpha\big|\nabla_{\mathbf{R}}\beta\rangle - \frac{1}{2M}\langle\alpha\big|\nabla_{\mathrm{e}}^2|\beta\rangle, \qquad \text{(S10)}$$

$\mathbf{F}_{\alpha\beta}$, according to the relation in Eq. (S6), is the antisymmetric derivative coupling vector,

$$\mathbf{F}_{\alpha\beta} = \frac{1}{\mu}\langle\alpha|\nabla_{\mathbf{R}}\beta\rangle, \qquad \text{(S11)}$$

and $V_\alpha$ is the BO potential,

$$V_\alpha = \langle\alpha|\hat{H}_{\mathrm{el}}|\alpha\rangle. \qquad \text{(S12)}$$

**Identical nuclear charges:** When both nuclei have the same charge, the electronic solutions to the clamped-nuclei Hamiltonian, Eq. (S4), are necessarily insensitive to the nuclear masses. Therefore, they exhibit a well-defined $g/u$ symmetry in this case. It is then advantageous to switch from the CNM frame to the center of nuclear charges (CNC) coordinates when calculating the matrix elements defined in Eqs. (S10) and (S11). The transformation from the CNM to the CNC system is expressed through the following relations:

$$\mathbf{R}(\mathrm{C}) = \mathbf{R}, \qquad \mathbf{r}_i(\mathrm{C}) = \mathbf{r}_i + \delta\mathbf{R}, \qquad \text{(S13)}$$

where

$$\delta = \frac{1}{2}\frac{M_A - M_B}{M}. \qquad \text{(S14)}$$

The derivatives with respect to the coordinates of the CNM system are thus

$$\nabla_{\mathbf{R}} = \nabla_{\mathbf{R}(\mathrm{C})} + \delta\sum_{i=1}^{N}\nabla_{\mathbf{r}_i(\mathrm{C})}, \qquad \nabla_{\mathbf{r}_i} = \nabla_{\mathbf{r}_i(\mathrm{C})}. \qquad \text{(S15)}$$

The operator $\nabla_{\mathrm{e}}$, defined in Eq. (S3), does not change its form in the CNC frame, so the operator $\nabla_{\mathbf{R}}$ may be written in a simpler form:

$$\nabla_{\mathbf{R}} = \nabla_{\mathbf{R}(\mathrm{C})} + \delta\nabla_{\mathrm{e}}. \qquad \text{(S16)}$$

Note that the first part of this operator, $\nabla_{\mathbf{R}(\mathrm{C})}$, has $g$ symmetry, while $\nabla_{\mathrm{e}}$ has $u$ symmetry.

The matrix elements $D_{\alpha\beta}$ from Eq. (S10) can now be split into two parts,

$$D_{\alpha\beta} = D_{\alpha\beta}^{g} + D_{\alpha\beta}^{u}, \qquad \text{(S17)}$$

where

$$D_{\alpha\beta}^{g} = \frac{1}{2\mu}\langle\nabla_{\mathbf{R}(\mathrm{C})}\alpha\big|\nabla_{\mathbf{R}(\mathrm{C})}\beta\rangle - \frac{1}{8\mu}\langle\alpha\big|\nabla_{\mathrm{e}}^2|\beta\rangle, \qquad D_{\alpha\beta}^{u} = \frac{\delta}{2\mu}\big(\langle\nabla_{\mathbf{R}(\mathrm{C})}\alpha|\nabla_{\mathrm{e}}\beta\rangle + \langle\nabla_{\mathrm{e}}\alpha|\nabla_{\mathbf{R}(\mathrm{C})}\beta\rangle\big). \qquad \text{(S18)}$$

For electronic functions of $\Sigma$ symmetry, the components of $\mathbf{F}_{\alpha\beta}$ [Eq. (S11)] perpendicular to the internuclear axis vanish by symmetry. Therefore, $\mathbf{F}_{\alpha\beta}$ can be further simplified to

$$\mathbf{F}_{\alpha\beta} = \mathbf{n}F_{\alpha\beta}^{g} + \mathbf{n}F_{\alpha\beta}^{u}, \qquad \text{(S19)}$$

where $\mathbf{n} = \mathbf{R}/R$ and

$$F_{\alpha\beta}^{g} = \frac{1}{\mu}\langle\alpha\big|\mathbf{n}\cdot\nabla_{\mathbf{R}(\mathrm{C})}\beta\rangle, \qquad F_{\alpha\beta}^{u} = \frac{\delta}{\mu}\langle\alpha\big|\mathbf{n}\cdot\nabla_{\mathrm{e}}\beta\rangle. \qquad \text{(S20)}$$

Let us now consider a specific two-state model [Eq. (S8)], where the electronic functions have $\Sigma_u^+$ symmetry ($\psi_1 = \psi_u$) and $\Sigma_g^+$ symmetry ($\psi_2 = \psi_g$). Additionally, the nuclear functions have the form $f_\alpha(\mathbf{R}) =$ $\frac{\chi_\alpha(R)}{R} Y_{JM}(\theta,\phi)$, where $Y_{JM}(\theta,\phi)$ are spherical harmonics. In this case the $2\times 2$ nuclear matrix operator from Eq. (S9) can be rewritten as

$$\mathbf{H}(R) = \begin{pmatrix} -\frac{1}{2\mu}\frac{\mathrm{d}^2}{\mathrm{d}R^2} + \frac{J(J+1)}{2\mu R^2} + D_{uu}^{g} + V_u & -\frac{1}{2}\left(F_{ug}^{u}\frac{\mathrm{d}}{\mathrm{d}R} + \frac{\mathrm{d}}{\mathrm{d}R}F_{ug}^{u}\right) + D_{ug}^{u} \\ \frac{1}{2}\left(F_{ug}^{u}\frac{\mathrm{d}}{\mathrm{d}R} + \frac{\mathrm{d}}{\mathrm{d}R}F_{ug}^{u}\right) + D_{ug}^{u} & -\frac{1}{2\mu}\frac{\mathrm{d}^2}{\mathrm{d}R^2} + \frac{J(J+1)}{2\mu R^2} + D_{gg}^{g} + V_g \end{pmatrix}. \qquad \text{(S21)}$$

The diagonal terms $D_{uu}^{g}$ and $D_{gg}^{g}$ are known as the diagonal BO corrections (DBOC) to the interaction potential[34,35]. For large internuclear distances, the off-diagonal term $F_{ug}^{u}$ vanishes while $D_{ug}^{u}$ reaches its asymptotic limit, which is equal to

$$-\frac{1}{4}\left(\frac{1}{M_A} - \frac{1}{M_B}\right)\left(\langle\nabla_\mathrm{e}^2\rangle_{N_1} - \langle\nabla_\mathrm{e}^2\rangle_{N_2}\right), \qquad \text{(S22)}$$

where $\langle\nabla_\mathrm{e}^2\rangle_{N_k}$ represents the expectation value of the operator $\nabla_\mathrm{e}^2$ calculated with $N_k$-electronic atomic wave function ($N_1 + N_2 = N$). Since the isotope shift of an atom in the first order of the electron-nucleus mass ratio is given by

$$\Delta E_{X,N_k} = -\frac{1}{2M_X}\langle\nabla_\mathrm{e}^2\rangle_{N_k}, \qquad \text{(S23)}$$

Eq. (S22) may be identified as half the difference between two separate sums of atomic isotope shifts. These sums correspond to the two possible dissociation limits, where the electrons are distributed between nuclei $A$ and $B$ as either $(N_1, N_2)$ or $(N_2, N_1)$.

**Potential energy curves:** We determine the PECs, $V_u$ and $V_g$, for the ground $X^2\Sigma_u^+$ and first excited $A^2\Sigma_g^+$ electronic states of the $He_2^+$ molecular ion utilizing the full configuration interaction (FCI) method implemented in the HECTOR code[36]. To achieve sub-wavenumber accuracy, the calculations of the BO potentials are performed using a family of correlation-consistent Gaussian-type orbital basis sets. The largest basis set employed in these calculations, characterized by the cardinal number $X = 9$, comprises 532 basis functions per atom. The final PECs are augmented with scalar relativistic corrections[22].

At large internuclear separations, both potentials are smoothly matched to the long-range interaction,

$$V_\mathrm{LR}(R) = -\frac{C_4}{R^4} - \frac{C_6}{R^6} - \frac{C_7}{R^7} - \frac{C_8}{R^8} - \frac{C_9}{R^9}, \qquad \text{(S24)}$$

with coefficients $C_4 = 0.6916$, $C_6 = 1.5970$, $C_7 = 3.6635$, $C_8 = 9.8177$, and $C_9 = 37.84$ in atomic units[22].

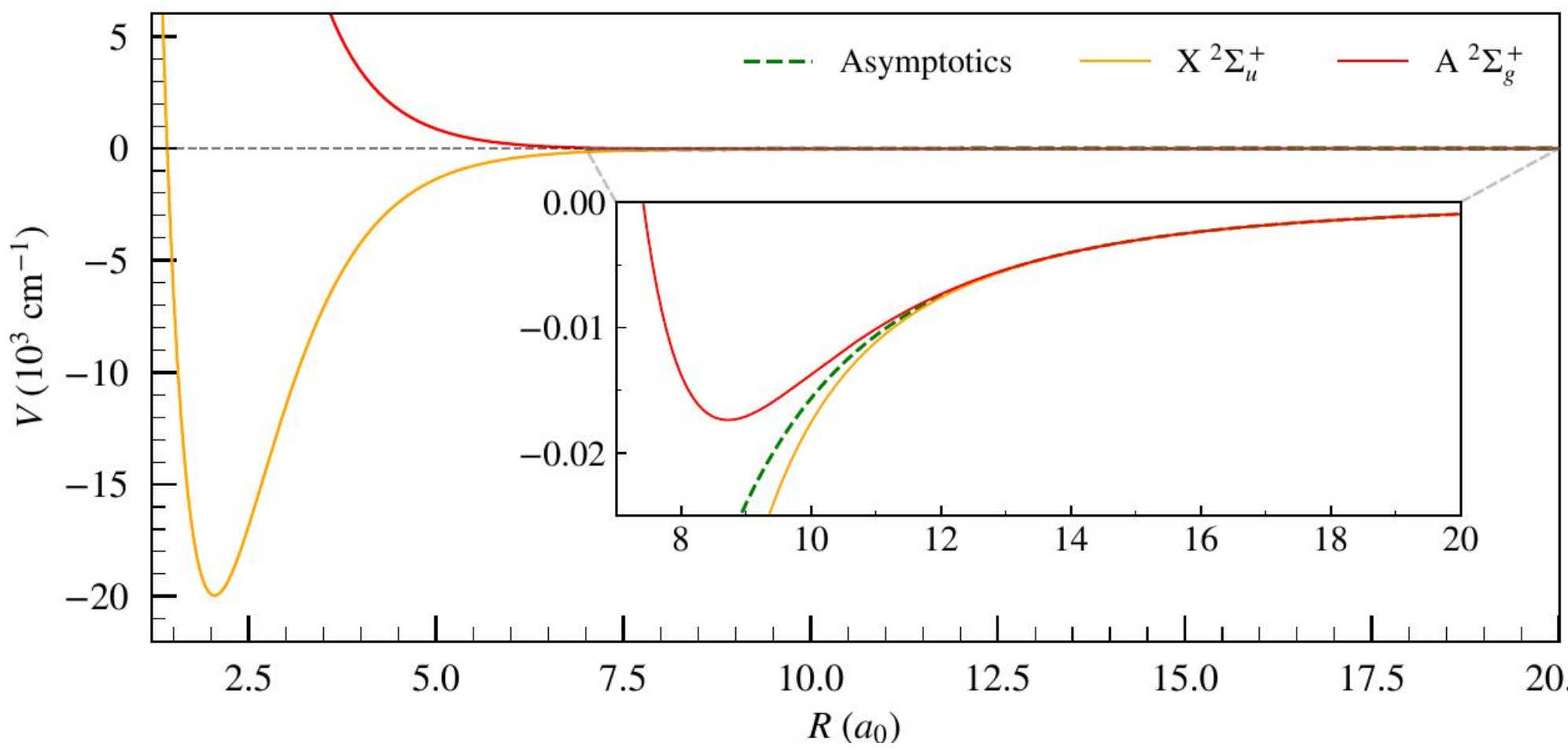

**Extended Data Fig. 1**: PECs for the $X^2\Sigma_u^+$ and $A^2\Sigma_g^+$ electronic states of the $He_2^+$ molecular ion. The plot illustrates the behavior of the potentials and their transition into the analytical long-range asymptotics (indicated by the green dashed line). The inset provides a magnified view of the shallow minimum of the $A^2\Sigma_g^+$ state.

**Reaction rate calculations**

**$^4He^+ + {}^4He$**

To calculate the reaction rate for the homonuclear case, we employ the method proposed by Massey and Smith[37]. For the $^4He^+/^4He$ system, where the total nuclear spin is $s = 0$, the differential cross section is given by

$$I_{\mathrm{tot}}(\theta) = \frac{1}{4}\left|f_g(\theta) + f_u(\theta) + f_g(\pi - \theta) - f_u(\pi - \theta)\right|^2 = \frac{1}{4}|f_{\mathrm{el}}(\theta) + f_{\mathrm{ct}}(\pi - \theta)|^2, \qquad \text{(S25)}$$

where $f_g(\theta)$ and $f_u(\theta)$ are the scattering amplitudes associated with the gerade and ungerade potential energy curves, respectively, and $f_{\mathrm{el}}(\theta)$ and $f_{\mathrm{ct}}(\pi - \theta)$ are the elastic and backward charge-exchange scattering amplitudes.

The scattering amplitudes on the gerade and ungerade potential energy curves are obtained from the partial-wave expansion

$$f_{g,u}(\theta) = \frac{1}{2ik}\sum_{\ell=0}^{\ell_{\max}}(2\ell+1)\left(e^{2i\delta_\ell^{g,u}} - 1\right)P_\ell(\cos\theta), \qquad \text{(S26)}$$

where $\delta_\ell^{g,u}$ are the phase shifts for the gerade and ungerade potentials, respectively, and $P_\ell(\cos\theta)$ are Legendre polynomials. The amplitudes at the exchange angle are obtained by evaluating $f_{g,u}(\pi - \theta)$.

The total cross section is obtained by integrating the differential cross section over the full angular range $0 \leq \theta \leq \pi$. We have also performed a partial-wave analysis to identify the dominant partial waves responsible for the resonance peaks.

**$^4He^+ + {}^3He$**

**Single Channel Propagation:** Along with the exact coupled-channel calculations, we also evaluate the $^4He^+$ + $^3He$ system under the assumption of a purely resonant charge-exchange process. In this approximation, the differential cross section is written as

$$I_{\mathrm{ch}}(\theta) = \frac{1}{4}\left|f_g(\theta) - f_u(\theta)\right|^2, \qquad \text{(S27)}$$

which corresponds to the standard resonant charge-exchange expression obtained from the difference of the gerade and ungerade scattering amplitudes. This simplified treatment neglects nonadiabatic coupling and assumes complete coherence between the two electronic states, as commonly employed in resonant charge-exchange and spin-exchange models. Here we have used the same two potential curves, but changed reduced mass. The propagation remains identical to the $^4He^+$ + $^4He$ calculations here. However, we see that this does not capture the rate coefficients well, and we need to proceed to coupled-channel propagation to capture the correct experimental measurements.

**Coupled Channel Propagation:** We use our own implementation of a coupled-channel solver named CoupledChan. We employ Johnson's log-derivative propagation method[38] with an adaptive step size based on the local wavelength. The solution is propagated from 1.2 bohr to 1000 bohr, with the local step adaptively set to one five-hundredth of the local wavelength. We obtain S-matrices for each partial wave with a given angular momentum and collision energy by matching the long-range boundary conditions in terms of modified spherical Bessel functions. From the partial wave S-matrices, we calculate partial elastic and inelastic cross sections, as well as differential cross sections.

To treat the non-adiabatic radial derivative coupling term in Eq. (S21), we employ the diabatization technique[39]. We introduce the mixing angle $\beta(R)$, which defines the unitary transformation from the adiabatic to the diabatic basis as a function of the internuclear distance $R$:

$$\mathbf{U}\big(\beta(R)\big) = \begin{pmatrix} \cos\big(\beta(R)\big) & -\sin\big(\beta(R)\big) \\ \sin\big(\beta(R)\big) & \cos\big(\beta(R)\big) \end{pmatrix}. \qquad \text{(S28)}$$

The function $\beta(R)$ is related to the non-adiabatic derivative coupling via the relation

$$\beta(R) = \beta_0 + \int_{r_0}^{R} \mu\, F_{ug}^{u}(R')\, \mathrm{d}R', \qquad \text{(S29)}$$

where $\beta_0$ and $r_0$ are arbitrary constants that do not affect the results. Transforming the Hamiltonian from Eq. (S21) to the diabatic representation yields

$$\mathbf{H}_{\mathrm{d}}(R) = \begin{pmatrix} -\frac{1}{2\mu}\frac{\mathrm{d}^2}{\mathrm{d}R^2} - \frac{\mu}{2}\left|F_{ug}^{u}\right|^2 & 0 \\ 0 & -\frac{1}{2\mu}\frac{\mathrm{d}^2}{\mathrm{d}R^2} - \frac{\mu}{2}\left|F_{ug}^{u}\right|^2 \end{pmatrix} + \mathbf{U}^{\dagger}\big(\beta(R)\big)\begin{pmatrix} V_u^{\mathrm{eff}} & D_{ug}^{u} \\ D_{ug}^{u} & V_g^{\mathrm{eff}} \end{pmatrix}\mathbf{U}\big(\beta(R)\big), \qquad \text{(S30)}$$

where $V_\alpha^{\mathrm{eff}}$ is defined as

$$V_\alpha^{\mathrm{eff}} = \frac{J(J+1)}{2\mu R^2} + D_{\alpha\alpha}^{g} + V_\alpha, \qquad \text{(S31)}$$

The unwanted non-adiabatic radial derivative terms are canceled by the action of the kinetic energy operator on $\mathbf{U}\big(\beta(R)\big)$, which additionally generates the scalar correction term $-\frac{\mu}{2}\left|F_{ug}^{u}\right|^2$.

The final coupled-channel calculations for $^4He^+ + {}^3He$ and $^3He^+ + {}^4He$ collisions are carried out using the complete Hamiltonian of Eq. (S30), including all non-adiabatic couplings and their radial dependence. However, compared with a model that includes only the asymptotic values of the non-adiabatic couplings, the inclusion of the $R$-dependence of the non-adiabatic derivative couplings and of the diagonal and off-diagonal BO corrections leads only to minor modifications of the reaction rates, well within the experimental error bars.